# Controlled protein synthesis and spatial organisation in microfluidic environments

Aukse Gaizauskaite[1], Emma E. Crean[1], Imre Banlaki[1], Jan L. Kalkowski[1], Henrike Niederholtmeyer[1, 2, *]

[1] Technical University of Munich (TUM), Campus Straubing for Biotechnology and Sustainability, Germany
[2] SynBioFoundry@TUM, Technical University of Munich, Straubing, Germany
[*] Corresponding author

## Abstract

Performing cell-free expression (CFE) in tailored microfluidic environments is a powerful tool to investigate the organisation of biosystems from molecular to multicellular scales. While cell-free transcription-translation systems simplify and open up cellular biochemistry for manipulation, microfluidics enables miniaturisation and precise control over geometries and reaction conditions. In this review, we highlight the benefits of combining microfluidics with CFE reactions for the study and engineering of molecular functions and the construction of life-like systems from non-living components. By defining spatial organisation at different scales and sustaining non-equilibrium conditions, microfluidic environments play a key role in the quest to "boot up" the biochemistry of life.

## Graphical abstract:

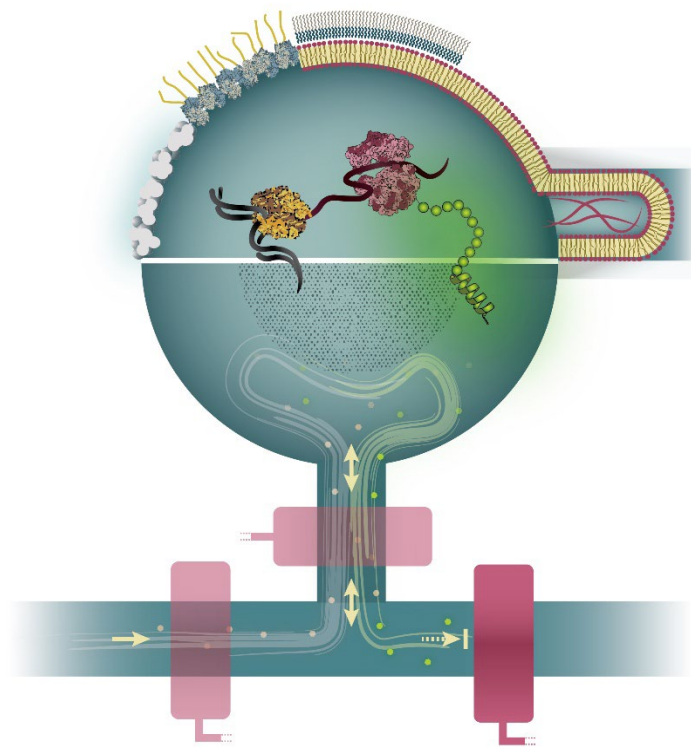

**Highlights:**

- Cell-free expression (CFE) in microfluidic environments serves as a powerful tool to investigate the organisation of biosystems from molecular to multicellular levels;
- Different shapes, geometries, and scales of microfluidic CFE setups simulate the conditions for accurate, cell-like regulation of biochemical processes;
- Microfluidic setups for CFE reactions enable (semi)continuous exchange of reagents and can maintain non-equilibrium steady states, which are a defining feature of living cells;
- The combination of CFE with microcompartments elevates automated high-throughput experiments and expands the borders of current knowledge about the spatial organisation of living systems.

**Keywords**: synthetic biology, cell-free expression, microfluidics, spatial organisation, artificial cells

**Introduction**

Cell-free expression (CFE) systems harness the cellular machinery to synthesise RNA and proteins in a simplified environment that is amenable to manipulation. The scalable volumes of CFE reactions and mix-and-match workflow synergise with automation and liquid handling tools. By increasing throughput and eliminating time-consuming steps such as plasmid cloning or protein purifications, CFE-based workflows have sped up "design-build-test-learn cycles" in synthetic biology and resulted in the optimisation of regulatory sequences, RNAs, proteins, and entire reaction networks. At each iteration through the engineering cycle, well-designed assays or enrichment schemes can lead to improved designs for the following rounds through rational design, *in vitro* evolution, or machine learning [1–6]. To enable screens for DNA-encoded functions, the link between genotype and phenotype is essential and is usually established by spatial separation between reactions [1–11]. Here, small-volume liquid handlers have advanced rapid cell-free prototyping in microtiter plates [1,6–8]. Taking miniaturisation a step further, droplet microfluidics generates water-in-oil emulsions that boost throughput even more (**Figure 1A, 1C**) [3,4,9–11].

Besides screening applications, CFE systems are a starting point for attempts to engineer a self-replicating synthetic cell and complex life-like systems capable of self-organisation [12,13]. With the goal of life-like systems in mind, some limitations of cell-free systems become obvious. The reactions are mostly performed in a closed batch format, where they rapidly approach chemical equilibrium, the equivalent of cell death. Another difficulty is the regulation of a large number of partly incompatible biochemical reactions to orchestrate metabolism, cell growth, and division. For an efficient integration of all the biochemical processes required for life, these processes likely need to be controlled in time and space. Indeed, compartmentalisation and spatial order are defining features of life: not only at the scale of a cell, but also at the molecular level or larger, multicellular scales. Because spatial organisation, non-equilibrium conditions, and tight regulation are so central to life, they are important engineering targets for the vision of bottom-up synthetic life.

Microfluidic tools have played an essential role in work on creating life-like systems from non-living matter, which we will highlight in this review. These tools are instrumental, as they operate at cellular scales, enable bespoke geometries, and allow precision handling of fluids via pico injection, micropumps, and microvalves [10,14]. We will showcase recent studies that take advantage of microfluidics to achieve compartmentalisation, adjust reaction conditions in real time and maintain a non-equilibrium state. This enables researchers to interrogate biochemical processes down to the single-molecule level, engineer self-organised patterns and mimic the conditions needed to sustain the biochemistry of life.

**Creating cell-like compartments**

Droplet microfluidics produces nanoliter to picoliter microreactors for massively parallel, miniaturised assays. Such spherical, micron-sized droplets commonly serve as compartments for boundary-enclosed cell-free reactions. Surfactant-stabilised water-in-oil droplets can be produced microfluidically in a high-throughput manner, yielding monodisperse synthetic reaction compartments (**Figure 1A**). They have been employed in directed evolution experiments with genotype–phenotype linkage [9], for engineering regulators [11], screening and characterising new peptide and protein variants (**Figure 1C**) [5,15,16], optimising biocircuits [17], and creating cell-free microreactors with multi-gene expression capabilities [18].

Monodisperse single and double emulsions have served as cell-like compartments to study numerous biochemical processes; however, they are hermetic, trapping any function inside. While this is desirable in most screening applications [15], it becomes a limitation when trying to build increasingly life-like systems. Such a system, capable of material exchange and communication with its environment, would ideally be contained by a functionalised lipid membrane. As a substitute membrane, proteinosomes or porous polymer capsules can be used to implement permeable boundaries capable of material exchange and communication, where selectivity of the compartment boundary is realised by size-exclusion (**Figure 1B**). Typically, genetic information is immobilised within such compartments, so material exchange is limited to the diffusion of metabolites and, for highly porous membranes, synthesised proteins [19,20]. In contrast to the crosslinked membranes of proteinosomes or porous polymer capsules, droplet interface bilayers establish semipermeable boundaries specialised for "cell-cell" contacts and communication (**Figure 1 E**). Closely arrayed single emulsion droplets create contact lipid-bilayers [21] through which they can exchange small signalling molecules and achieve collective behaviour such as quorum sensing [22]. Lipid membranes, in the form of polydisperse vesicles or liposomes, have seen wide use in combination with CFE to create simple, minimal cell mimics [12,13,23]. In contrast to bulk production, microfluidic generation of monodisperse liposomes, via double emulsions, has enabled the fabrication of cell-like compartments with tailored sizes and internal cargo [10,24,25].

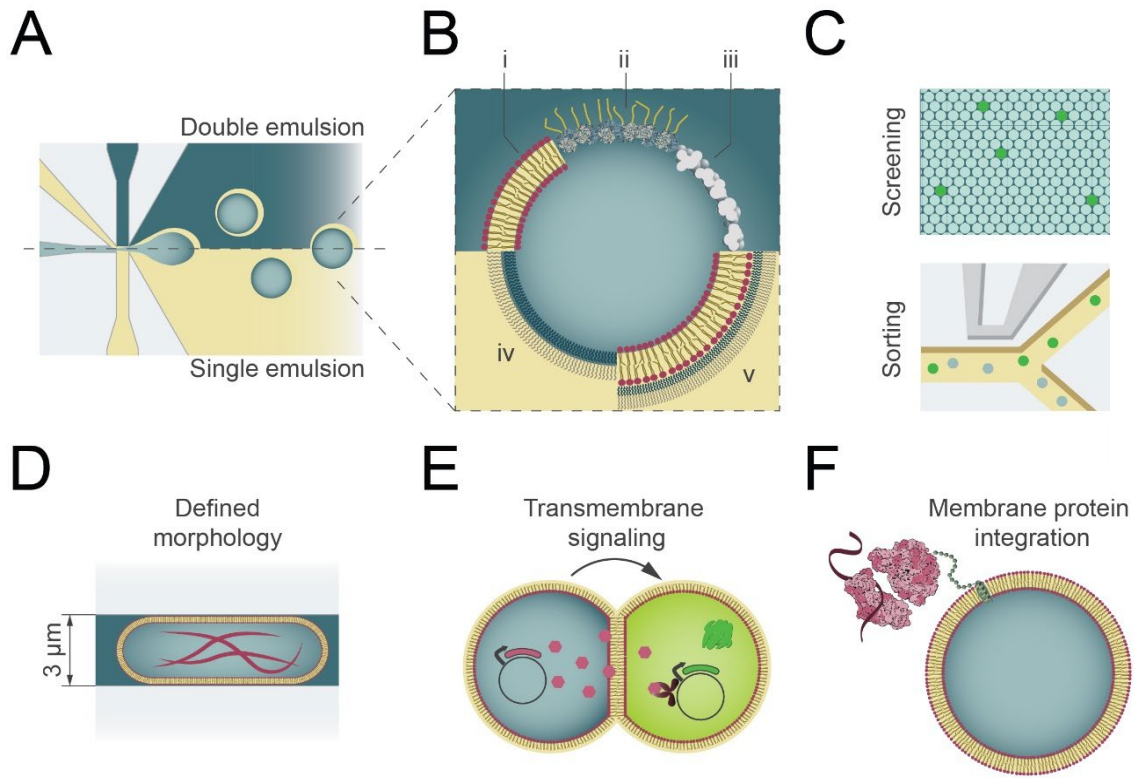

**Figure 1**. Cell-like microcompartments and droplet microfluidics. **A)** Flow focusing chips can create monodisperse water-in-oil (bottom) or water-in-oil-in-water (top) emulsions with tunable sizes. Droplet composition, such as encapsulated cargo and amphiphiles at the interfaces, is adjustable [9,15]. **B)** Different compartment boundaries and interfaces enable tailor-made interactions and barrier functions. **(i)** Lipid bilayers simulate a cell-like interface [25]. Porous membranes with tunable permeability can be created from crosslinked protein-conjugates **(ii)** [19] or synthetic polymers **(iii)** [20]; **(iv)** Single emulsions are stabilised by lipids or synthetic surfactants [16]; **(v)** a supported lipid bilayer can be assembled at the interface of a surfactant-stabilised single emulsion to create a biocompatible boundary [10]. **C)** Compartments can be evaluated by screening for CFE activity in a microscope (top) [3] or fluorescence-activated droplet sorting (bottom) [11]. **D)** Deforming CFE-loaded liposomes in microfluidic chips creates defined shapes to influence internal assembly processes [46]. **E)** CFE-filled single emulsions can be brought into contact to form interfacial lipid bilayers and to allow communication across the membrane [22]. **F)** Membrane proteins can be integrated into lipid bilayers during translation without the need for surfactants [23].

In addition to a boundary, a synthetic cell needs to maintain a multitude of biochemical functions. One important function is the expansion of its membrane from soluble precursors, which may be realised by producing natural or unnatural enzyme pathways for phospholipid synthesis in liposomes [23,26]. CFE enables the functionalisation of membranes via cotranslational integration of transmembrane proteins into lipid bilayers [27,28]. Direct cotranslational integration, without the help of surfactants, can directionally insert membrane proteins [23,29] and probe hydrophobic mismatch between proteins and membranes (**Figure 1F**) [28].

Furthermore, minimal liposome systems are suitable to investigate cell division [12], to synthesise genetically encoded artificial cytoskeleton structures [30] and implement Darwinian evolution of self-replicating DNA [13]. Increasingly complex spatial organisation was achieved with *in situ* produced biomolecular condensates [31], creating a compartment inside a compartment architecture [32]. Finally, even higher structural organisation can be attained to mimic and study simple proto-tissues [33,34], and communication within artificial tissue structures [35].

**Creating cellular scales and shapes**

Scale and geometry have a significant impact on biochemical reactions and assembly processes [12,36]. Deliberate chip design not only reduces the use of reagents but enables access to microscale physics such as boundary effects, mass transport, and stochasticity, more accurately emulating the conditions within a living cell. Controlling the scale of the microsized compartments enables the study of biological systems at cellular dimensions [25,37–41], single molecule quantification in a defined environment [42], as well as the study of stochastic events [37,43]. Control of scale achieved by varying chip geometry can be used to rationally prepare droplets of defined sizes [44].

While unconstrained droplets take on a spherical shape, accurate reconstitution of many cellular processes — including those involving cytoskeletal dynamics, molecular assemblies, or reaction-diffusion systems — requires non-spherical compartments, or other defined geometries [36,45,46]. Droplets can be deformed into non-spherical shapes, either by adjusting the material composition of the compartment [47] or by using a microfluidic platform for physical confinement (**Figure 1D**) [46]. Work from the Dekker group demonstrated the importance of compartment morphology by deforming droplets into rod shapes, similar to those adopted by many microbes. In the controlled confinement, tubulin fibres preferentially oriented in alignment, while spherical compartments displayed no preferential orientation [46], showing that the shape of the compartment is critical for the arrangement of protein networks.

**Expanding control through microfluidics**

Precise control of reaction conditions is essential for the construction of increasingly complex and life-like biochemical systems. While select studies have been able to achieve a high degree of compositional control, the closed batch reaction format limits possibilities to simultaneously alter and monitor reaction conditions [48]. At increasing sample numbers, the successive dispensing and measuring of kinetics in plates becomes difficult, even if assisted by liquid handlers or high-throughput droplet generation systems.

In contrast, confinement of reactions in microfluidic chambers maintains accessibility for direct manipulation while affording large numbers of compartments from micron- to nano-scales [14,49]. The experiment can be initiated or perturbed at any point by introducing reagents through fluid channels and integrated valves [50,51], even during ongoing measurements (**Figure 2A**). The merits of such dynamic experiments were demonstrated by the high-throughput engineering of complex, dynamic circuits in microfluidic chambers based on continuous, steady-state cell-free protein expression (**Figure 2B**) [14]. In another study, precisely timed additions of fuel enabled repetitive control of growth, division, and re-growth of artificial, phase-separated compartments in a microfluidic chamber [52].

To create communicating arrays of cell-like compartments, chambers can be interconnected. For example, work of the Bar-Ziv group demonstrated the formation of reaction-diffusion patterns from coupled genetic oscillators (**Figure 2C**). Communication between interconnected microfluidic chambers led to synchronised oscillations and patterns reminiscent of multicellular collective behaviour [53,54]. While microfabricated connections between chambers precisely define the speed and routes of molecular exchange, boundary-free transcription and translation in microfluidic devices harbouring immobilised DNA templates offer a number of different opportunities. For example, heterogeneous surface chemistry separated synthesis reactions from the assembly of the small ribosomal subunit to reveal hierarchical and cooperative interactions between ribosomal RNAs and proteins [55]. Another example of boundary-free synthesis repurposes Illumina chips by deliberately stalling transcription and translation to co-localise the synthesised protein with its DNA template [56]. This technique combined protein assays and DNA sequencing in a single device to achieve high-affinity antibody discovery [2].

Real-time measurements during protein synthesis provide insights into folding and assembly processes, for example, to quantify affinities between binding partners while concentrations of synthesised proteins increase [57]. To elucidate protein folding, CFE allows site-specific incorporation of fluorescently labelled amino acids for single-molecule fluorescence resonance energy transfer (FRET) observations during translation and folding [58]. Unstable, difficult-to-purify proteins could be synthesised, tethered, and analysed by single-molecule force spectroscopy within microfluidic flow cells to investigate protein folding [59]. Other custom devices can measure dynamic interactions under non-equilibrium conditions, such as Min protein oscillations and association of phage defence protein on tethered membranes [60], or a gene regulatory circuit on a single immobilised DNA molecule for localisation [61].

**Maintenance of non-equilibrium conditions**

Microfluidic chips can be designed to enable semi-continuous or continuous exchange of reagents. The influx of fresh reagents, accompanied by the efflux of products, results in a non-equilibrium steady state, which can be maintained for extended periods of time. This contrasts with traditional batch reactions, which occur within a closed system and reach equilibrium relatively quickly. A non-equilibrium system offers distinct benefits, chiefly the ability to conduct experiments over extended time scales and to showcase dynamic systems. Crucially, it also mimics the inherent non-equilibrium state of living cells, a property that is a defining feature of life.

Diffusion of small molecules across a permeable membrane is one way to achieve this. On-chip phospholipid (**Figure 2D**) [62] or hydrogel [63] membranes, which separate a reaction chamber from a feed solution, permit small molecule exchange while retaining larger proteins. Continuous diffusion of small molecules can extend protein synthesis to over 24 hours and improve yields by 7-fold [62,63]. Expanding beyond the diffusion of small molecules, microfluidic chips also enable the exchange of larger macromolecules and DNA for periodic partial exchange of the solution in the reaction chamber (**Figure 2A**) [64]. The study of dynamic gene networks, such as oscillators, is impossible in batch reactions due to the accumulation of products and the lack of metabolic homeostasis. Several studies have highlighted the versatility of semi-continuous and continuous flow microfluidics to capture dynamics that would not be observable in closed batch reactions (**Figure 2 A-C**) [53,54,63–66]. In such microfluidic environments, the rates of exchange tune steady state protein levels and dynamics like oscillation periods [63,64,66]. Similarly, non-equilibrium conditions by controlled diffusive exchange were essential to generate dynamic patterns from coupled oscillators in space (**Figure 2C**) [53,54].

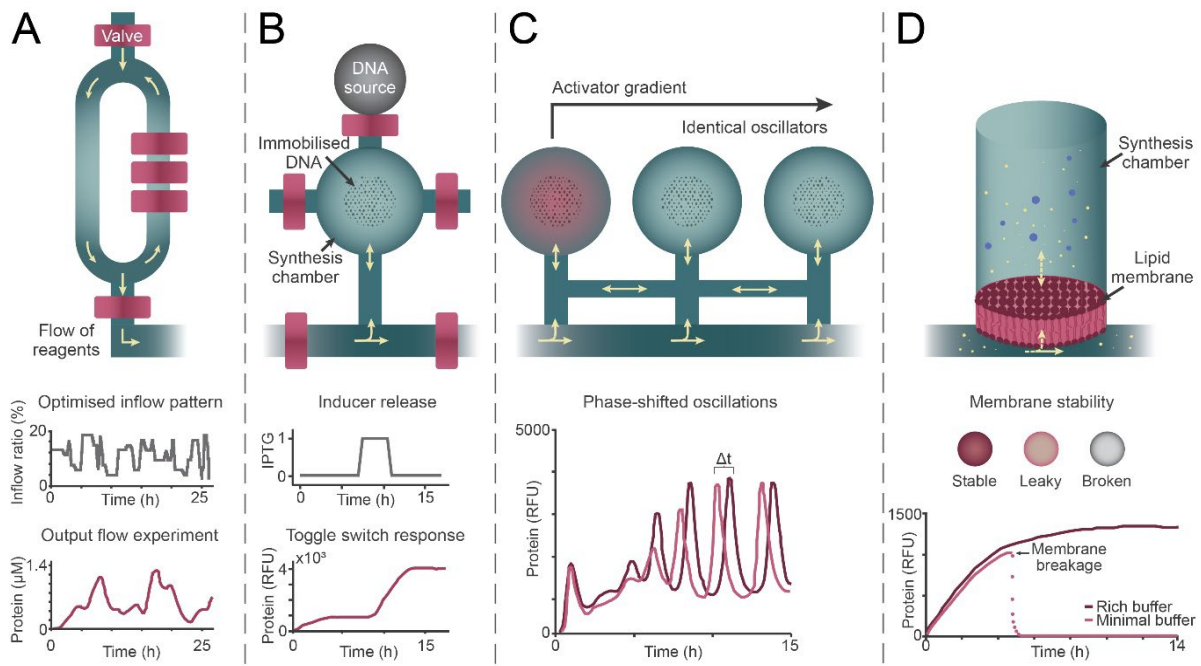

**Figure 2**. Examples for microfluidic control of cell-free expression reactions. **A)** Ring-shaped micro-reactor with semi-continuous flow of CFE reagents, controlled by micro-valves. Bottom graphs show reporter protein synthesis output in response to a changing inflow of reagents [67]. **B)** Chamber with micro-valves ensuring a steady-state CFE reaction environment for dynamic genetic circuits. Graphs show toggle switch response upon inducer exposure [14]. **C)** Interconnected microchambers with diffusion channels for communication between genetic circuits. The graph shows phase-shifted oscillations observed in synthesis chambers exposed to an activator gradient [53]. **D)** Picoliter-scale microwell sealed by a phospholipid membrane that allows passive nutrient supply. The graph shows CFE activity and membrane stability for a nutrient-rich buffer and a minimal buffer [62]. Examples using free DNA: **A, D;** immobilised DNA: **B, C**. Red rectangles in **A** and **B**: positions of pneumatic microvalves.

The ability to exchange DNA, CFE machinery and reaction products in time creates flexibility in experimental designs. For example, the ability to input triggers for different durations, at different time points, and concentrations provides granular data on network characteristics, enabling detailed modelling and simulations (**Figure 2A**) [66,67]. In another example, programmable exchange of DNA templates and the transcription-translation machinery allowed Lavickova et al. to demonstrate a CFE system that is able to partially self-regenerate [68]. Regeneration of the protein synthesis machinery is a vital step towards constructing self-sustaining synthetic cells.

## Conclusions and summary

The integration of continuously evolving CFE systems with microscale environments is driving the field of engineering biology. From closed reaction confinements to continuous synthesis in chips, microfluidic CFE reactions advance in throughput, miniaturisation, and precision. This achieves faster screening for new functions, linking genotype and phenotype, and controlling biochemical processes in real-time.

Importantly, microfluidic CFE reactions can be considered a stepping stone for bottom-up reconstruction of complex systems and ultimately synthetic life. The defining features of life, such as spatial organisation and precise maintenance of non-equilibrium conditions, have been extremely challenging to replicate *in vitro*. Fortunately, the transition from biomolecule synthesis to highly ordered functional assemblies is now enabled by combining microfluidic environments and cell-free synthesis, bringing these systems closer towards fully functional synthetic cells. Current efforts are establishing cell-like architectures that replicate the meticulous regulation of biological processes at cellular, subcellular and multicellular scales.

In the coming years, completely automated high-throughput experiments integrated with artificial intelligence will revolutionise the development of designer biomolecules and expand the borders of current knowledge about the self-organisation of living systems. We anticipate that CFE, in combination with microfluidics, will enable this vision and yield key contributions to the field of engineering and bottom-up synthetic cell biology.

## Acknowledgements:

The authors acknowledge funding by Deutsche Forschungsgemeinschaft (DFG, German Research Foundation) through grant NI 2040/1-1 and by the European Research Council through ERC Starting Grant SYNSEMBL (101078028).

## Declaration:

The authors declare no conflict of interest.

## Author Contributions:

**AG:** Conceptualisation, Visualisation, Writing - original draft, Writing - review and editing. **EEC:** Conceptualisation, Writing - original draft, Writing - review and editing. **IB:** Conceptualisation, Visualisation, Writing - original draft, Writing - review and editing. **JLK:** Conceptualisation, Writing - original draft, Writing - review and editing. **HN:** Conceptualisation, Funding acquisition, Supervision, Writing - review and editing. EEC, IB and JLK contributed equally to this review. Their positioning within the author list was determined by rolling a die for initiative.

# References and recommended reading:

Papers of particular interest, published within the period of review, have been highlighted as:

\* of special interest,

\*\* of outstanding interest.

Cell-free expression assays explore interactions between *de novo* designed membrane proteins of varying hydrophobic thicknesses and synthetic membranes. Hydrophobic mismatch between the protein and the membrane leads to spatial segregation of proteins into different compartments or membrane domains.

This publication demonstrates sequential enzyme synthesis and enzyme activity measurements in high-throughput parallel microfluidic devices. Each device is able to analyse over 1500 enzyme variants in parallel.

This publication generates large-scale (5mm x 5mm) 2D oscillation patterns on a microfluidic chip. By modifying the geometries of the microfluidic chip, the authors can alter the coupling strength and therefore the associated dynamics of the oscillator.

This study takes advantage of the gradually increasing protein concentrations during CFE reactions to measure binding curves of protein-protein interactions by fluorescence correlation spectroscopy, and demonstrates a proof-of-concept for measurements in microfluidically produced double emulsion droplets.